\documentclass[aps,prl,showpacs,preprintnumbers,twocolumn,superscriptaddress]{revtex4-1}
\usepackage{amsmath,amssymb}
\usepackage{bm}
\usepackage{tipa}
\usepackage{upgreek}
\usepackage{comment}
\usepackage{mathrsfs}
\usepackage{graphicx}
\usepackage{braket}
\usepackage{enumitem}
\usepackage{mathbbol}
\usepackage{gensymb}
\usepackage[normalem]{ulem}
\usepackage{color}
\usepackage[dvipsnames]{xcolor}
\usepackage[colorlinks,bookmarks=true,citecolor=blue,linkcolor=red,urlcolor=blue]{hyperref}
\usepackage{hyperref}

\begin{document}

\title{Disorder-free localization transition in a two-dimensional lattice gauge theory}

\author{Nilotpal Chakraborty}
\thanks{Corresponding author}
\email{nilotpal@pks.mpg.de}
\affiliation{Max-Planck-Institut f\"{u}r Physik komplexer Systeme, N\"{o}thnitzer Stra\ss e 38, Dresden 01187, Germany}

\author{Markus Heyl}
\affiliation{Max-Planck-Institut f\"{u}r Physik komplexer Systeme, N\"{o}thnitzer Stra\ss e 38, Dresden 01187, Germany}
\affiliation{Theoretical Physics III, Center for Electronic Correlations and Magnetism, Institute of Physics, University of Augsburg, D-86135 Augsburg, Germany}

\author{Petr Karpov}
\affiliation{Max-Planck-Institut f\"{u}r Physik komplexer Systeme, N\"{o}thnitzer Stra\ss e 38, Dresden 01187, Germany}

\author{Roderich Moessner}
\affiliation{Max-Planck-Institut f\"{u}r Physik komplexer Systeme, N\"{o}thnitzer Stra\ss e 38, Dresden 01187, Germany}

\begin{abstract}
Disorder-free localization represents a novel mechanism for ergodicity breaking in lattice gauge theories (LGTs) which can even occur in two spatial dimensions (2D). It has been shown that the U(1) quantum link model (QLM) can localize due to an emergent classical percolation transition fragmenting the system into disconnected real-space clusters. While the nature of the quantum localization transition (QLT) is still debated for conventional many-body localization, here we provide the first comprehensive characterization of the QLT for the QLM in 2D for a disorder-free case.
In this work we find compelling evidence that the QLT in the 2D QLM is continuous and we determine its universality class. We base our considerations on a spectral analysis of finite-size clusters in the percolation problem which exhibits two regimes - one in which large clusters effectively behave non-ergodically, a result naturally accounted for as an interference phenomenon in configuration space, and the other in which all large clusters behave ergodically.Our analysis can also be applied to other 2D U(1) LGTs potentially including also matter degrees of freedom.

\end{abstract}

\maketitle

The quantum localization transition (QLT) has been of great interest since Anderson and co-workers' discovery that an electron hopping on a one- or two-dimensional lattice can localize in presence of disorder \cite{AndersonLoc,abrahams1979scaling}. Pioneering work has found such a transition even in presence of weak interactions implying the presence of an extended nonergodic so called many body localized (MBL) phase \cite{basko2006metal}. While the MBL phase is believed to be stable in one dimension (1D) \cite{imbrie2016many}, the opposite has been argued in two dimensions (2D) \cite{DeRoekaval,PotirnicheAva}. However, the nature of the QLT in 1D is still debated and it is unclear if it is continuous such as the noninteracting Anderson transition~\cite{oganesyanPRB,PalMBL,PotirnicheAva,DeRoekaval,Prosenchaos,morningstar2021avalanches,abanin2021distinguishing}. 

Recent works show that localization also occurs in models of lattice gauge theories (LGTs) without quenched disorder - a phenomenon dubbed disorder free localization (DFL) \cite{Smith2017,Brenes2018}. Localization in these LGTs is due to the emergence of so-called gauge superselection sectors, which induce an effective internal quenched disorder for the dynamics. Most models for DFL are in 1D, where quantum interference induces localization similar to conventional Anderson localization \cite{Smith2017,Brenes2018,IreneDFL,halimeh2021enhancing,halimeh2021stabilizing,LiPolar,HodsonDFL}. Recently, certain 2D LGTs have also shown to exhibit localization due to the presence of a non-percolating phase in an emerging classical percolation problem describing the gauge charges \cite{KarpovPRL,Robertovar}.
However, the characterization of the resulting disorder-free QLT has remained elusive.
Settling the nature of the transition is particularly demanding in interacting many-body systems, such as these LGTs, where a treatment beyond perturbative arguments is necessary.

One central result of this work is the precise characterization of the QLT in the 2D U(1) quantum link model (QLM) on a square lattice, involving an accurate estimate of the quantum critical point and its associated universality class including critical exponents. Besides the above central result, we also make the following advances: i) Despite the lattice being fragmented into disconnected clusters in the underlying classical percolation problem, we find evidence that quantum interference can, in addition, play a crucial role in the dynamics. We show via level-spacing statistics that the larger clusters emerging in the vicinity of the classical percolation transition, and on the non-percolating side, experience a crossover from ergodic to non-ergodic on tuning the ratio of the potential and kinetic couplings. This result implies that the long-time dynamics on the ergodic side of the QLT will eventually thermalize only for very large systems and very long times. Similar claims of non-ergodicity in experimentally relevant sample sizes with a slow drift towards ergodicity have been made in a recent work about the MBL regime \cite{morningstar2021avalanches}. ii)Another important result of our work is a concrete prediction on the dynamical properties of observables close to the QLT. We show that for a space- and time-dependent spin correlator the asymptotic long-time value reflects critical exponents of classical percolation.

\begin{figure}[t]
\centering
\includegraphics[width = 1.0 \columnwidth, height = 3.5cm]{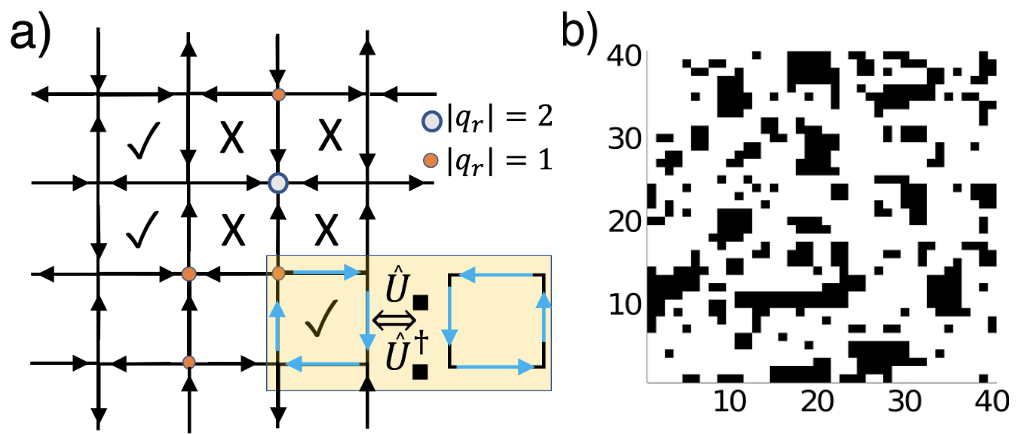}
\caption{a) Sample initial configuration for Monte Carlo flip process, $\uparrow$ and $\rightarrow$ represent spin $\ket{\uparrow}$ whereas $\downarrow$ and $\leftarrow$ represent spin $\ket{\downarrow}$. The ticked plaquettes are flippable due to their clockwise/anticlockwise orientation, background charges are indicated by the circles on the sites. The crossed plaquettes are completely frozen due to charge 2 sites. The box indicates the flip process where the $\hat{U}_{\square}$ operator flips all the spins (coloured blue) on plaquette P. b) Disconnected clusters of flipped plaquettes (in black) in the non-percolating phase for the state $\ket{\psi(\alpha = 0)}$ on a $40\times40$ lattice. Frozen plaquettes are shown in white.}
\label{fig:perc1}
\end{figure}

\textit{2D U(1) QLM.---} The QLM is a discrete version of a Wilsonian LGT where instead of continuous U(1) parallel transporters, the degrees of freedom on links are discrete spins \cite{Wiese2013,chandrasekharan1997quantum}. Due to its discrete nature the QLM is amenable to numerics and experiment \cite{Wiese2013}. We consider the following spin-$1/2$ Hamiltonian with a plaquette flip term (kinetic energy) with coupling constant $J$ and a Rokhsar-Kivelson potential \cite{RokhsarPRL} with coupling constant $\lambda$, which counts the number of flippable plaquettes:
\begin{equation}
    \hat{H} = \lambda \sum_{\square} (\hat{U}_{\square} + \hat{U}_{\square}^{\dagger})^2 - J \sum_{\square} (\hat{U}_{\square} + \hat{U}_{\square}^{\dagger}) \, ,
    \label{eq:HamiltonianQLM}
\end{equation}
where $\hat{U}_{\square} = \hat{S}^{+}_{\textbf{r},i} \hat{S}^{+}_{\textbf{r}+\textbf{i},j} \hat{S}^{-}_{\textbf{r}+\textbf{j},i} \hat{S}^{-}_{\textbf{r},j}$ flips all the spins on a plaquette if they are oriented clockwise or anticlockwise. If this condition is met, we say that the plaquette is flippable; for an illustration see Fig.~\ref{fig:perc1}a. Here, $\hat{S}^{+/-}_{\textbf{r},\mu}$ denotes the spin raising/lowering operators for the spin joining site $\textbf{r}$ and $\textbf{r} + \hat{\mu}$, where $\hat{\mu} = \hat{i}/\hat{j}$ is the lattice vector. 
$\hat{H}$ in Eq.~(\ref{eq:HamiltonianQLM}) satisfies local constraints due to the U(1) gauge symmetry, generators of which have the form $\hat{G}_{\mathbf{r}}= \sum_{\hat{\mu}} (\hat{S}^z_{\mathbf{r},\hat{\mu}} - \hat{S}^z_{\mathbf{r} - \hat{\mu},\hat{\mu}})$ and can be interpreted as the net electric field flowing into the site. These generators commute with $\hat{H}$ and with each other on different sites, hence eigenvalues of $G_{\mathbf{r}}$, $q_{\mathbf{r}} \in \{-2,-1,0,1,2\}$, can label eigenstates of $\hat{H}$. The set $\mathbf{q} = \{q_{\mathbf{r}_1},q_{\mathbf{r}_2,...}\}$ for all lattice sites $\mathbf{r}_i$ defines a superselection sector in Hilbert space with states satisfying $G_{\mathbf{r}} \ket{\psi (\mathbf{q})} = q_{\mathbf{r}} \ket{\psi (\mathbf{q})} $. These states are kinematically disconnected from states with different $\mathbf{q}$. Due to the net electric field interpretation of $G_{\mathbf{r}}$, $q_{\mathbf{r}}$ can be interpreted as a static background charge satisfying Gauss' law. Such discrete link models have been widely studied in condensed matter theory, with examples ranging from the toric code \cite{kitaev2003fault} and quantum spin ice \cite{Hermele2004,ShannonPRB} to quantum dimer models \cite{RokhsarPRL,Moessnerising}. Experimental proposals to realize such models in cold atom settings have also been made \cite{Glaetzle2014,CeliPRX} with a recent proposal implemented for finite sizes in a 2D Rydberg atom setup \cite{semeghini2021probing}, however, their non-equilibrium dynamics \cite{Debashishscars} are still widely unknown due to the numerical challenges associated with treating 2D interacting quantum matter .

\begin{figure}
\centering
    \includegraphics[width = 8cm, height = 4.6cm]{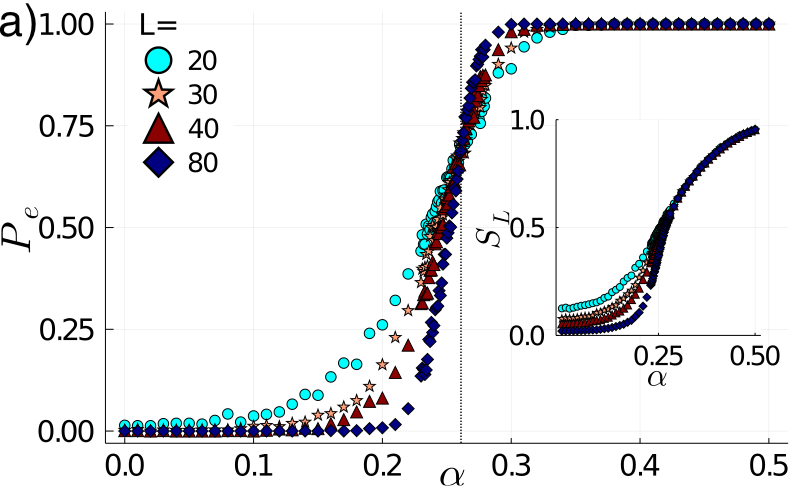}
    \includegraphics[width = 8.1cm, height = 5cm]{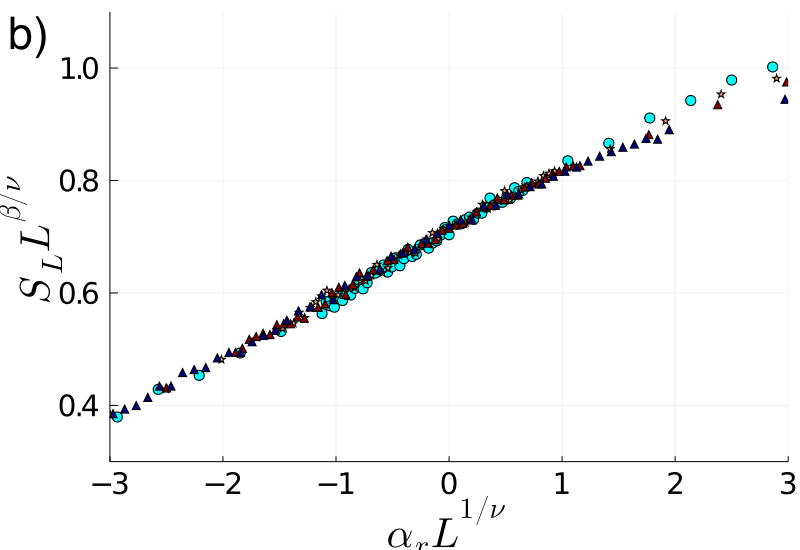}
    \caption{Results for the percolation problem on square lattices of linear dimension $L = 20, 30, 40$ and $80$. We average over 1000 sectors for each $\alpha$ a) Probability that a cluster wraps around either the x or the y axes vs $\alpha$. The dotted line represents the percolation threshold $\alpha_c$ (Inset) Fraction of plaquettes occupied by the largest cluster vs $\alpha$ b) Scaling collapse for the order parameter with the critical exponents $\alpha_c = 0.261$, $\nu = 4/3$ and $\beta = 5/36$. Here $\alpha_r = (\alpha - \alpha_c)/\alpha_c$}
    \label{fig:pprob}
\end{figure}

\textit{Bounding the QLT through a classical percolation problem.---}
DFL in LGTs arises due to an effective disorder average due to background charges in typical superselection sectors \cite{Brenes2018}. To characterize the QLT we choose the following parameterizaiton for the initial state which allows us to access different superselection sectors \cite{KarpovPRL}.
\begin{equation}
    \ket{\psi_0 (\alpha)} = \otimes_i \big[ \text{sin}(\alpha + \pi/4)\ket{\mathrm{FF}_i} + \text{cos}(\alpha + \pi/4)\ket{\overline{\mathrm{FF}}_i} \big]
    \label{initstate}
\end{equation}
where $\alpha \in  [0, \pi/4]$ controls the density of background charges. $\ket{\mathrm{FF}_i}$ and $\ket{\overline{\mathrm{FF}}_i}$ denote the orientation of the $i^{\text{th}}$ spin in two possible fully flippable states (all plaquettes flippable -- checkerboard configurations of clockwise and anticlockwise plaquettes). For $\alpha = 0$ we get the product state where each spin can be $\ket{\uparrow}$ or $\ket{\downarrow}$ with equal probability, such a state is distributed over all super-selection sectors. For $\alpha = \pi / 4$ we get the product state from the fully flippable, i.e. zero background charge sector. LGTs like in Eq.~(\ref{eq:HamiltonianQLM}) with finite local Hilbert spaces can be analyzed via a classical correlated percolation problem \cite{KarpovPRL}, which we summarize for completeness. For every $\alpha$ in Eq.~(\ref{initstate}) we get an initial state distributed over certain superselection sectors. One can choose a random basis state (sector) from this distribution, perform an infinite temperature Monte Carlo (MC) search for a plaquette and then flip it if flippable. We then define a site percolation problem on a square lattice, where every plaquette is considered occupied if flippable. Clusters in the problem are defined as a set of connected plaquettes, each of which has been flipped more than 100 times (this defines the exit criteria for the MC search; if this criteria is not met we exit after $10^8$ searches, we have checked for convergence on varying number of searches). Static background charges $q_{\mathbf{r}}$ are unchanged throughout the plaquette flip process. We study the percolation transition in system sizes of up to linear dimension $L = 80$. To determine the percolation threshold we go beyond Ref.~\cite{KarpovPRL} and calculate the percolation probability using wrapping probabilities \cite{Newmannperc}, in which percolation occurs if there exists a cluster which wraps onto itself. This can occur in topologically distinct ways and in Fig.~\ref{fig:pprob}a we calculate the probability that a cluster wraps around either in the $x$ or in the $y$-direction. This approach is better compared to percolation probabilities calculated for open boundary conditions as in \cite{KarpovPRL} since the former has O($1/L^2$) finite size corrections as compared to O($1/L$) of the latter. Ref.\cite{Newmannperc} also showed that $\alpha_c$ estimated from finite size wrapping probabilities converges to the critical value for the infinite system as $|\alpha_c(L)-\alpha_c| \sim L^{-11/4}$. From Fig.~\ref{fig:pprob}a we infer that for a  finite parameter regime in $\alpha$ we get a non-percolating phase comprising disconnected clusters (Fig.~\ref{fig:perc1}b), in agreement with Ref \cite{KarpovPRL}.Using our improved percolation model we go beyond the analysis in ref \cite{KarpovPRL} and determine the nature of the transition via a scaling collapse of the order parameter $S_L$ (inset of Fig.~\ref{fig:pprob}a) defined as the fraction of plaquettes occupied by the largest cluster, using the value for $\alpha_c = 0.261$  . We choose critical exponents $\nu = 4/3$ and $\beta = 5/36$, which gives a good collapse as in Fig.~\ref{fig:pprob}(b). Hence, our percolation problem is in the universality class of the standard 2D site percolation problem with short-range or no spatial correlations. In the non-percolating phase it is clear that real-space fragmentation due to background charges induces localization. Such a classical threshold sets a bound for the QLT, the percolation algorithm, however, is insufficient to determine the properties of the quantum transition and  the exact quantum transition point which depend on quantum dynamics of finite-size clusters. There could be further localization within larger clusters,which would make the quantum and classical transitions differ.

\begin{figure}
\includegraphics[scale = 0.35]{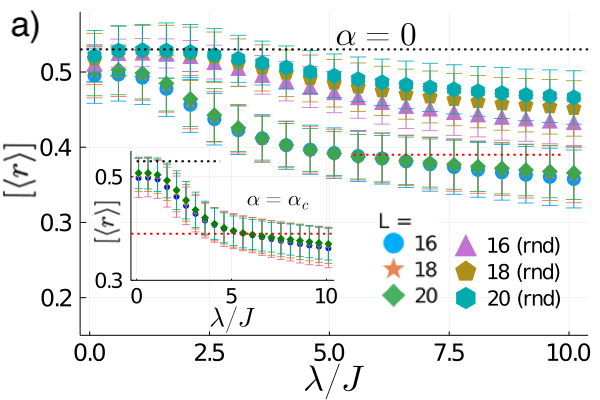}
\includegraphics[scale = 0.33]{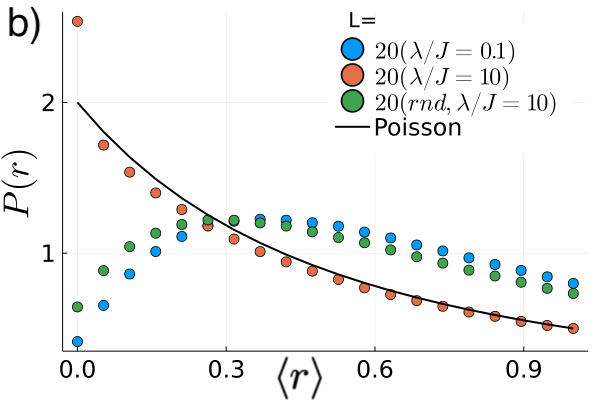}
\caption{a) The gap ratio averaged over the spectrum and over all clusters with 16, 18 and 20 plaquettes in $N = 1500$ superselection sectors of the $80\times80$ lattice  vs $\lambda/J$ for state $\ket{\psi(\alpha = 0)}$. Here, ``rnd'' implies results for the randomized energy model with same configuration space connectivity. Error bars indicate variation due to different cluster shapes, (inset) same data for $\ket{\psi(\alpha = \alpha_c)}$ with $N = 2500$. b) Corresponding probability distribution for a 20 plaquette cluster.} 
\label{fig:gapratio}
\end{figure}

\begin{figure}
\includegraphics[width=1.0\linewidth,height = 8 cm]{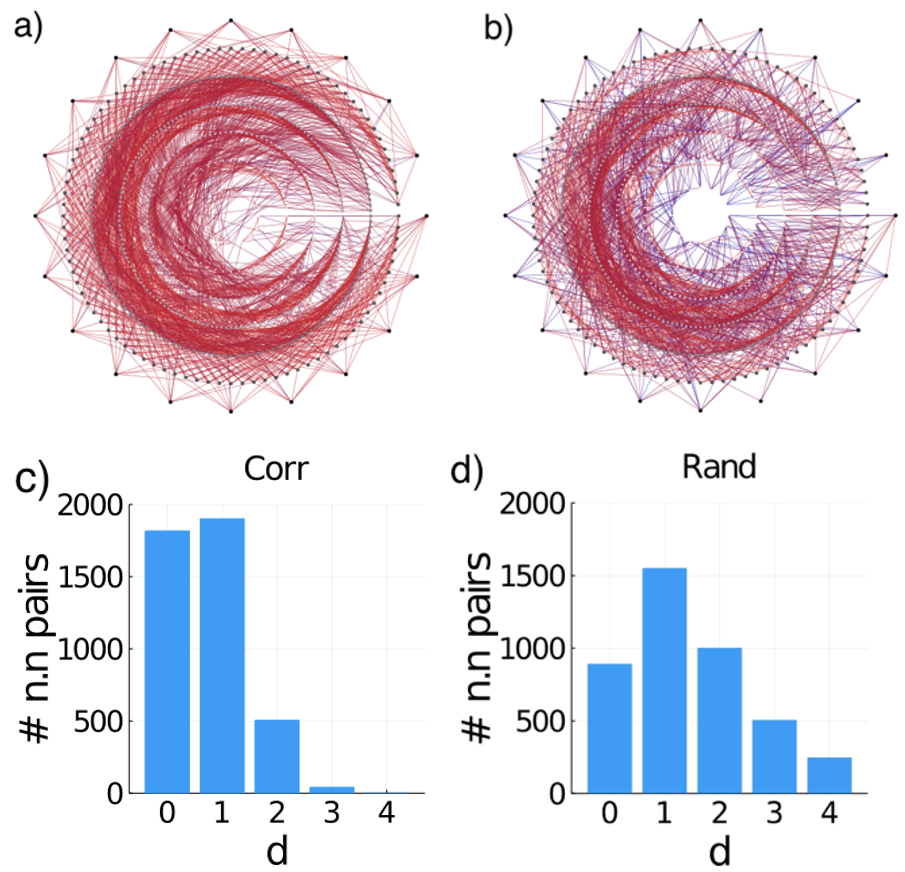}
\caption{Configuration space graph for a 16 plaquette cluster. (a,b) Graph with correlated and randomized on-site energies respectively. The edge colour goes from red to blue as separation between connecting shells increases. (c,d) Histogram for normalized difference of the on-site energies   $d=|E_i-E_j|/\lambda$ between nearest neighbours $i,j$ in (a) and (b) respectively.}
\label{fig:configspace}
\end{figure}

\textit{Quantum-classical coincidence and}  non-ergodicity in finite clusters due to quantum interference.---
In the percolation problem localization arises classically from removing bonds due to static background charges which render certain plaquettes unflippable. However, plaquettes in each cluster are kinetically connected through subsequent plaquette flips. For every $\alpha$ we get a certain cluster distribution in our classical percolation problem with a non-percolating phase for $\alpha < \alpha_c$, and at a fixed $\alpha$, $\lambda/J$ controls the quantum dynamics of finite clusters (Fig.~\ref{fig:gapratio}a) We find that finite size clusters for every $\alpha < \alpha_c$, experience a crossover from ergodic to non-ergodic behaviour upon tuning $\lambda/J$ in Eq.~(\ref{eq:HamiltonianQLM}). We demonstrate this by calculating adjacent gap ratios, a common diagnostic for distinguishing between ergodic and non-ergodic behaviour \cite{oganesyanPRB}. The adjacent gap ratio is defined as $r_n = \text{min}\{\delta_n,\delta_{n+1}\}/\text{max}\{\delta_n,\delta_{n+1}\}$ where $\delta_n = E_n - E_{n-1}$, $E_n$ is the energy of $n^{\text{th}}$ eigenstate and the eigenstates are arranged in ascending order of their energies. We average $r_n$ over the entire spectrum of eigenstates of the cluster and over all clusters of same size in $1500$ superselection sectors (disorder realizations) of the $80 \times 80$ lattice to get $[\langle r \rangle]$. For small $\lambda / J$, $\langle r \rangle$ approaches the mean of the Gaussian Orthogonal Ensemble (GOE) distribution ($\langle r \rangle_\mathrm{GOE}\approx 0.53$) and the probability distribution is GOE-like, as expected in a thermal phase. However, in the opposite limit, $\langle r \rangle$ approaches the mean of Poisson distribution ($\langle r \rangle_\mathrm{POI}\approx 0.39$) as we go to larger clusters and the distribution is Poissonian, as seen in Fig.~\ref{fig:gapratio}a and b. There is a drift in the gap ratio on increasing cluster size implying that the infinite percolating cluster might not show Poissionian statistics and this non-ergodicity is a finite-size effect. There is also a slight variation depending on cluster shape as shown by the error bars in Fig.~\ref{fig:gapratio}a. Since there is no real-space fragmentation within a cluster, i.e all plaquettes are kinematically connected,  we interpret the nonergodicity as a quantum interference effect.
We confirm that ergodicity for small $\lambda/J$ in finite-size clusters also persists at the critical point $\alpha_c$ of the percolation problem (inset of Fig.~\ref{fig:gapratio}a). Ergodicity in all large clusters for small $\lambda/J$ at $\alpha_c$ implies that in the small potential limit, information is spread out in all large clusters and due to the absence of any localization within large finite clusters the classical and quantum phase transitions coincide. Our finding of ergodic behavior for individual clusters allows us to precisely characterize the disorder free QLT in 2D, which is our central result. We conclude that, for small $\lambda/J$, the disorder-free QLT is a \emph{continuous} transition with critical exponents $\nu = 4/3$ and $\beta = 5/36$ and hence it belongs to same universality class as conventional percolation. 

\textit{Scaling of correlation functions near the QLT.---} 
Our findings on the QLT also allow for direct predictions on dynamical universal behavior of the U(1) QLM.
An essential property of localized systems is absence of transport of energy or spin. To quantify transport we consider the following connected correlation function $A_{\mathbf{r}}(t)$ and its long-time average $A_{\mathbf{r}}^\infty$:
\begin{align}
    A_{\mathbf{r}}(t)&=\langle \hat{S}_0^z (0) \hat{S}_{\mathbf{r}}^z (t) \rangle - \langle \hat{S}_0^z (0)\rangle \langle\hat{S}_{\mathbf{r}}^z (t) \rangle\\
    A_{\mathbf{r}}^{\infty} &=\lim_{T \rightarrow \infty} \frac{1}{T}\int_0^{T} dt A_{\mathbf{r}}(t)    \, .
\end{align}

Since $\alpha_c$ coincides with the QLT point for small $\lambda/J$, in this limit we can estimate scaling of the dominant contribution to $A_r^\infty$ near the transition using critical exponents obtained in Fig.~\ref{fig:pprob}. Close to the QLT, on the ergodic side ($\alpha > \alpha_c$), real space lattice fragments into an infinite cluster, some dynamic finite-size clusters (with characteristic sizes up to correlation length $\xi\sim|\alpha-\alpha_c|^{-\nu}$ of the standard 2D percolation universality class \cite{stauffer2018introduction}), and some inert plaquettes (which are not flipped even once). $A^{\infty}_{\mathbf{r}}$ can be affected only by clusters of diameter $\geq r$. For $r\gg\xi$, finite cluster contributions are exponentially suppressed, hence if we first take the limit $r\rightarrow\infty$ and then $\alpha \rightarrow \alpha_c$ only the infinite cluster contributes to $A^{\infty}_{\mathbf{r}}$.
From conventional percolation theory we know that the probability that a given spin belongs to the infinite cluster $P \sim (\alpha - \alpha_c)^\beta$, hence 
\begin{equation}
   A_{r\rightarrow\infty}^{\infty} \sim P^2 \sim  (\alpha - \alpha_c)^{2\beta},
   \label{eq:A_scaling}
\end{equation}
for $\alpha\geq\alpha_c$ and is zero for $\alpha<\alpha_c$. Note that order of limits is important here: first we took $T\rightarrow\infty$, then we take $r\rightarrow\infty$, only after which we can take $\alpha\rightarrow\alpha_c$ limit.

\textit{Configuration space mapping.--- }
As a next step, we show that the nonergodicity in finite clusters observed in Fig.~\ref{fig:gapratio} can be interpreted as a quantum interference effect in configuration space as done in MBL \cite{basko2006metal,Royfock,Macefock}. We map the Hamiltonian of a cluster as defined in Eq.~(\ref{eq:HamiltonianQLM}) (now we take the sum in Eq.~(\ref{eq:HamiltonianQLM}) to be over all plaquettes in the cluster) to a tight-binding model in configuration space with
\begin{equation}
    \hat{H}_{CS} = \sum_{i} E_i \ket{\hat{\mu}_i}\bra{\hat{\mu}_i} - J \sum_{\braket{ij}} \ket{\hat{\mu}_i}\bra{\hat{\mu}_j}
\end{equation}
where $\ket{\hat{\mu}_i}$ is a string of $-1$'s, 1's and 0's indicating the configuration of constituent plaquettes(+/- 1 for clockwise/anticlockwise orientation and 0 otherwise), $E_i = m \lambda$, where $m$ is the sum of absolute values of all string elements and $J$ is the nearest-neighbour hopping amplitude connecting configurations differing by a plaquette flip. Nearest neighbours $i$ and $j$ satisfy $|E_i - E_j| = d \lambda$ where $d \in \{0,1,2,3,4\}$, depending on cluster geometry. Hence, we get a single-particle problem in a heterogeneous graph with discrete correlated disorder. If we treat a pair of nearest-neighbours as a two-level system, eigenstates will be localized when $|E_i - E_j| \gg J$ and in resonance when $|E_i - E_j| \ll J$. Since $|E_i - E_j| \leq 4\lambda$, when $\lambda/J \ll 1$ every pair is resonant and the system is delocalized and ergodic. However, when   $\lambda/J \gg 1$, resonances drastically reduce since only pairs with same on-site energy are resonant and most pairs satisfy $|E_i - E_j| \gg J$ favoring non-ergodicity. 

In Fig.~\ref{fig:configspace}(a) we show the configuration space graph of a typical cluster with 16 plaquettes from the state $\ket{\psi(\alpha = 0)}$. The nodes are arranged in equi-energy shells and $E_i$ increases outwards. To highlight the importance of on-site energy correlations we randomly shuffle the energies while keeping same connectivity. On doing so, we lose signatures of non-ergodicity as shown in Fig.~\ref{fig:gapratio}, the gap ratio and distribution no longer resemble Poissonian behaviour. We speculate that this can be understood by comparing configuration space graphs in Fig.~\ref{fig:configspace}(a) and (b). For correlated disorder most edges connect nodes with same classical energy or with energies differing by $\lambda$, which is not the case for the randomized energy model as illustrated by the histograms in Figs.~\ref{fig:configspace}(c) and (d). Hence, correlations in configuration space, which are wiped out by random reshuffling, are central for nonergodicity at $\lambda/J\gg 1$. Therefore, localization in this graph is caused by quantum interference due to these correlations.


\textit{Summary and discussion.---}Exact critical exponents of a quantum transition in interacting 2D systems are notoriously hard to find. In MBL, the nature of the transition is still unclear and exact critical exponents are unknown \cite{Prosenchaos,abanin2021distinguishing}. We have provided a detailed characterization of the QLT including its critical exponents in the 2D U(1) QLM. We have also shown the presence of an experimentally relevant finite size non-ergodic regime in disorder free 2D systems.
A further key question is the influence of matter fields, which introduce a dynamic charge in these 2D LGTs, thereby inducing a coupling between the previously disconnected superselection sectors of the pure gauge theory.
3D generalizations of this work
leads to questions of localization of emergent photons in quantum spin-ice,  of interest as a strongly-coupled LGT \cite{Pacefine}.

\textit{Data availability.---}The data for the figures is openly available on Zenodo at \cite{Chakdata}.

\textit{Acknowledgements.---}
This project has received funding from the European Research Council (ERC) under the European Union’s Horizon 2020 research and innovation programme (Grant Agreement No. 853443), and M.H. further acknowledges support by the Deutsche Forschungsgemeinschaft via the Gottfried Wilhelm Leibniz Prize program. This work was in part supported by the Deutsche Forschungsgemeinschaft under grants SFB 1143 (project-id 247310070) and the cluster of excellence ct.qmat (EXC 2147, project-id 390858490). 


\end{document}